\documentclass[conference]{IEEEtran}
\usepackage{cite}
\usepackage{amsmath,amssymb,amsfonts}
\usepackage{algorithmic}
\usepackage{graphicx}
\usepackage{textcomp}
\usepackage{xcolor}
\usepackage{comment}
\usepackage{caption,subcaption}
\def\BibTeX{{\rm B\kern-.05em{\sc i\kern-.025em b}\kern-.08em
    T\kern-.1667em\lower.7ex\hbox{E}\kern-.125emX}}
\begin{document}

\title{Link Prediction using Graph Neural Networks \\ for Master Data Management}

\author{\IEEEauthorblockN{Balaji Ganesan}
\IEEEauthorblockA{
\textit{IBM Research, India}\\
bganesa1@in.ibm.com
}\\
\IEEEauthorblockN{Gayatri Mishra}
\IEEEauthorblockA{
\textit{IBM Data and AI}\\
gayamish@in.ibm.com
}
\and
\IEEEauthorblockN{Srinivasa Parkala}
\IEEEauthorblockA{
\textit{IBM Data and AI}\\
shsriniv@in.ibm.com
}\\
\IEEEauthorblockN{Matheen Ahmed Pasha}
\IEEEauthorblockA{
\textit{IBM Data and AI}\\
matpasha@in.ibm.com
}
\and
\IEEEauthorblockN{Neeraj R Singh}
\IEEEauthorblockA{
\textit{IBM Data and AI}\\
sneeraj@in.ibm.com
}\\
\IEEEauthorblockN{Hima Patel}
\IEEEauthorblockA{
\textit{IBM Research, India}\\
himapatel@in.ibm.com
}
\and
\IEEEauthorblockN{Sumit Bhatia}
\IEEEauthorblockA{
\textit{IBM Research, India}\\
sumitbhatia@in.ibm.com
}\\
\IEEEauthorblockN{Somashekhar Naganna}
\IEEEauthorblockA{
\textit{IBM Data and AI}\\
soma.shekar@in.ibm.com
}
}

\maketitle

\begin{abstract}
Learning graph representations of n-ary relational data has a number of real world applications like anti-money laundering, fraud detection, and customer due diligence. Contact tracing of COVID19 positive persons could also be posed as a Link Prediction problem. Predicting links between people using Graph Neural Networks requires careful ethical and privacy considerations than in domains where GNNs have typically been applied so far. We introduce novel methods for anonymizing data, model training, explainability and verification for Link Prediction in Master Data Management, and discuss our results.
\end{abstract}


\section{Introduction}

Relational Data consists of tuples where each tuple is a set of attribute/value pairs. A set of tuples that all share the same attributes is a relation. Such relational data can be presented in a table, json arrays among other forms.

In Master Data Management (MDM), one or more tuples in relational data can be resolved to an entity. Typically, an entity in this setting is a person or an organization, but there can be other types too. These entities may share explicit and implicit relations between them. A common way to represent these entities and relationships is a graph, where each entity is a node, and the relationships are links (edges) between the nodes.

Master Data Management includes tasks like Entity Resolution, Entity Matching, Non-Obvious Relation Extraction. While the enterprise customer data is predominantly stored as relational data, graph stores are widely used for visualization and analytics.

Link Prediction is the task of finding missing links in a graph. These links could be typed or untyped. Given a graph with several nodes and links, a model can be trained to learn embeddings of the nodes and links, and predict missing links in the graph. In recent years, Graph Neural Networks (GNN) are being used for link prediction and node classification tasks.

\begin{figure}
  \includegraphics[height=4cm, width=\columnwidth]{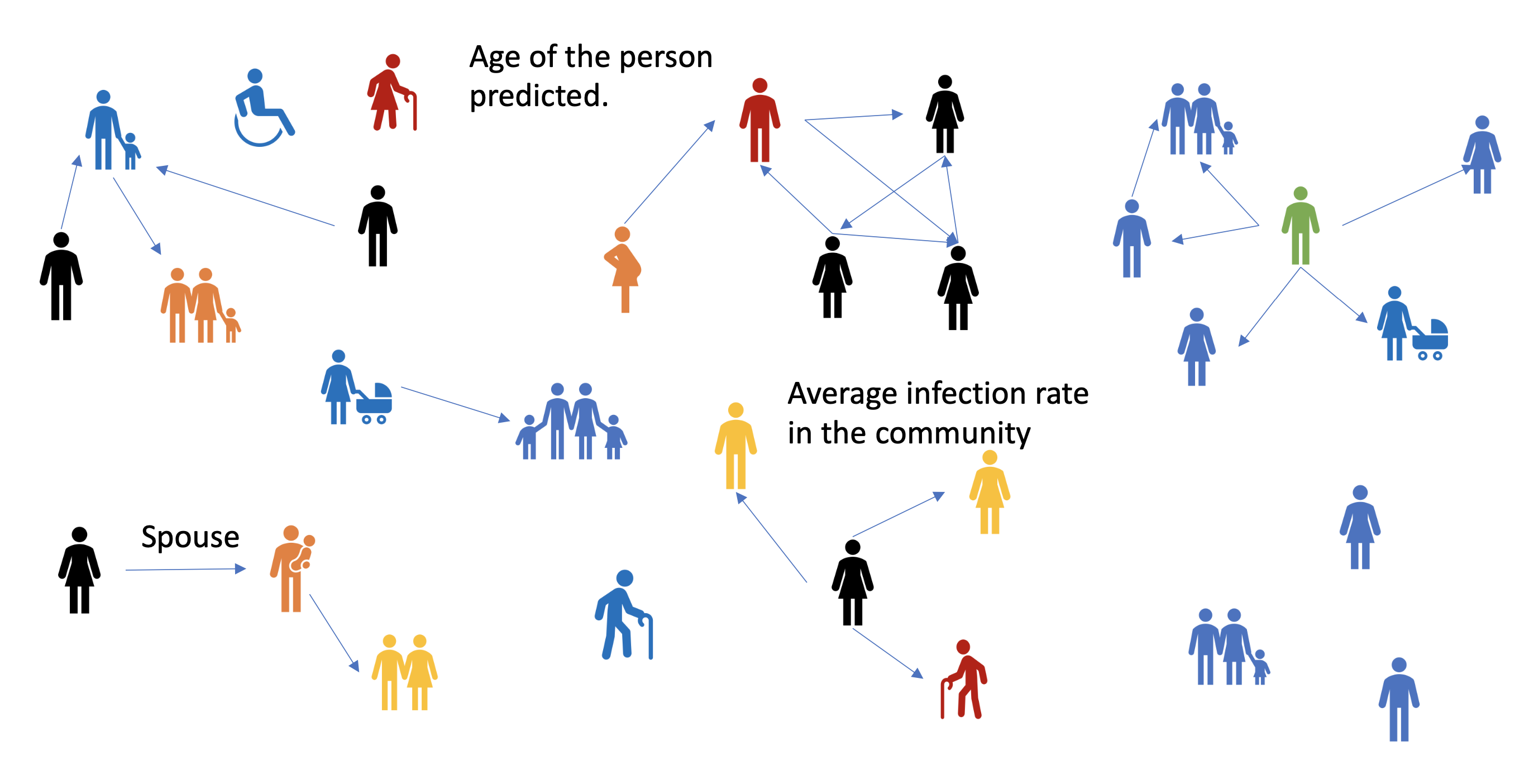}
  \caption{Contact tracing of COVID19 positive persons could be posed as a watchlist link prediction problem in MDM}
  \label{fig:watchlist}
\end{figure}

\begin{figure*}[htb]
\centering
    \includegraphics[width=\linewidth]{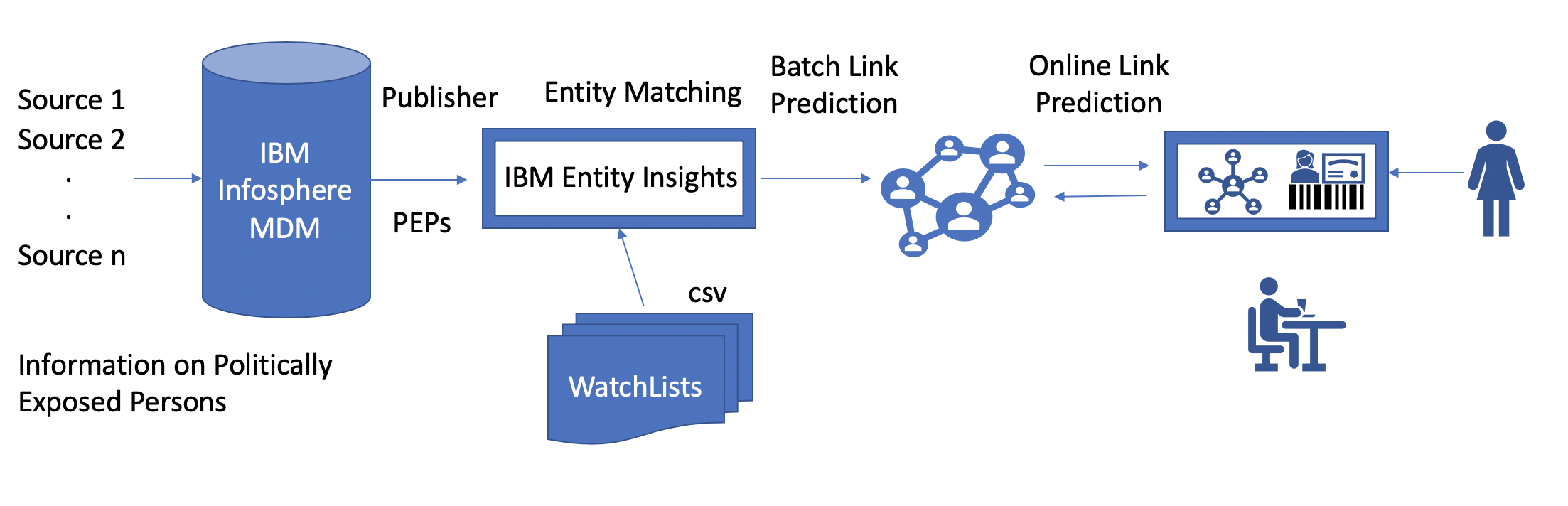}
    \caption{Predicting Links to Watch-List Nodes. A list of COVID19 persons can be uploaded as Watch-List and their links to people in a master data could be predicted. Note: We need informed consent of people for this contact tracing.}
    \label{fig:watchlist_link_prediction}
\end{figure*}

Graphs in MDM could be considered as Property Graphs which differ from Knowledge Graphs and Social Networks. In particular, MDM is focused on entities like people, organization, and location, which constitute nodes in the graph. Other entities like numerical ids, demographic information, business terms constitute the attributes of the nodes. Links between people or organizations in the real world constitute the edges of the graph. The type of people to people relations form the link type and details of the relationship like duration constitute the edge properties.

Link Prediction on people graphs presents a unique set of challenges in terms of model performance, data availability, fairness, privacy, and data protection. Further, Link Prediction among people has a number of societal implications irrespective of the use-cases. While link prediction of financial fraud detection, law enforcement, and advertisement go through ethical scrutiny, the recent use of social networks for political targeting and job opportunities also require ethical awareness on the part of the model developers and system designers.

Watchlist is a use-case typical in Master Data Management. Given a set of nodes in a graph, the task is to find links to Watchlist nodes from other nodes of the graph. For example, we can assume people who have tested positive for COVID19 as people on a Watchlist. We may want to find people who may have potentially come in contact with them, commonly known as contact tracing. This can however be a time consuming and potentially controversial process that impacts privacy.

In this work we discuss link prediction on property graphs involving people, their specific dataset requirements, ethical considerations, and practical insights in designing the infrastructure for industrial scale deployment.

The main contributions of this work are:

\begin{itemize}
\item We discuss the dataset requirements and ethical considerations to train neural models for many real-world applications in Master Data Management.
\item We present our results on training Graph Neural Networks on property graphs typical of MDM workloads.
\item We present 3 easily understandable explainability solutions in addition to interpretability, to explain the GNN model predictions.
\item We present GraphSheets, a method to help increase accountability in the development and deployment of our solution.
\end{itemize}

\section{Related Work}

\cite{schlichtkrull2018modeling} showed that a Relational Graph Convolutional Network can outperform direct optimization of the factorization (ex: DistMult). They used an autoencoder model consisting of an encoder – an R-GCN producing latent feature representations of entities and a decoder – a tensor factorization model exploiting these representations to predict labeled edges.

\cite{hamilton2017inductive} introduced GraphSAGE (SAmple and AggreGatE) an inductive framework that leverages node feature information (ex: text attributes, node degrees) to efficiently generate node embeddings for previously unseen data or entirely new (sub)graphs. In this inductive framework, we learn a function that generates embeddings by sampling and aggregating features from a node’s local neighborhood. \cite{you2019position} proposed a Position Aware Graph Neural Network that significantly improves performance on the Link Prediction task over the Graph Convolutional Networks.

\cite{guan2019link} introduced the WikiPeople dataset based on Wikidata. However, Wikidata does not have contact details and can be incomplete like the DBPedia (UDBMS) dataset. \cite{dasgupta2018hyte} also introduced a dataset based on Wikidata. Again, for afore mentioned reasons, it’s not different from DBPedia dataset. \cite{Zitnik2017} introduced the Protein-Protein Interaction dataset which has been used in a number of recent works in graph neural networks. [cite] presented Graph Convolutional networks. \cite{hamilton2017inductive} introduced the idea of inductive representation of nodes in a graph. \cite{you2019position} added anchor nodes to improve the representation of nodes in the graph.

\cite{oberhofer2014beyond} introduced many aspects of entity resolution using a probabilistic matching engine. DeepMatcher \cite{mudgal2018deep} presents a neural model for Entity Matching. \cite{konda2019executing} presented an end-to-end use case for Entity Matching. \cite{mullerintegrated} described an integrated neural model to find non-obvious relations.

The Protein-Protein Interaction (PPI) dataset introduced by \cite{Zitnik2017} consists of 24 human tissues and hence has 24 subgraphs of roughly 2400 nodes each and their edges. Having similar subgraphs helps to average the performance of the model across subgraphs.

Open Graph Benchmark \cite{ogb_benchmark} initiative by SNAP Group at Stanford is trying to come up with large benchmark datasets for research in Graph Neural Models.

Much of the recent work on explanations are based on post-hoc models that try to approximate the prediction of complex models using interpretable models. \cite{vannur2020data} present post-hoc explanations of the links predicted by a Graph Neural Network by treating it as a classification problem. They present explanations using LIME \cite{ribeiro2016should} and SHAP \cite{lundberg2017unified}. \cite{arya2020ai} introduced the AIX360 toolkit which has a number of explainability solutions that can be used for post-hoc explanation of graph models, if they can posed as approximated as interpretable models.

Over the years, Attention has been understood to provide an important way to explain the workings of neural models, particularly in the field of NLP. \cite{jain2019attention} challenged this understanding by showing that the assumptions for accepting Attention as explanation do not hold. \cite{wiegreffe2019attention} however argued that Attention also contributions to explainability.

More recently, \cite{agarwal2020neural} introduced Neural Additive Models which learn a model for each feature to increase the interpretability. In this work, we focus on solutions that are more suited to Graph Neural Models and in particular, explaining Link Prediction models.

\section{Ethical considerations in MDM Link Prediction}
\label{ethical_considerations}

Predicting links between people and organizations in the Master Data of any company, requires lot more care than innocuous use-cases like predicting links in protein-protein-interactions. In this section we describe a number of properties and ethical considerations that are specific to Link Prediction in Master Data Management.

\subsection*{Explainability}

Explainability methods in Graph Neural Networks tend to follow similar methods in text and images, namely identifying features that are most significant for the predictions. However, in enterprise applications, there is a need for explanations for non-technical users. Path-based explanations that help to visualize the neighborhood of the predicted links will require training data with medium sized and multiple paths between different people, as is common in the real world. The popular ‘six degrees of separation’ motivates having an average path length of 6 between nodes of a graph.

\subsection*{Verifiability}

In addition to Explainability, it is desirable to have additional data that can be used by Data Stewards to verify the predicted links. This idea is similar to the efforts in fact verification task. A piece of text that strengthens a predicted link, a heuristic measure like Jaccard coefficient, a measure of similarity (or lack thereof) between nodes involved in a link can all be used to verify the predicted links.

\subsection*{Temporal Information}

\cite{dasgupta2018hyte} introduced a new dataset for their work on temporal graph embeddings. Use-cases in financial fraud prevention require a dataset where the demographic attributes of a person change, sometimes quite drastically. Even otherwise, changes to attributes of people nodes is quite common in the real world (people moving residences, jobs, change in marital status or partners). Hence we’ll propose a dataset that captures events in the life of people.

\subsection*{Provenance and Lineage}

A closely related capability in enterprise graphs is the ability to maintain lineage and provide provenance of nodes, attributes and relations. For example, if the graph were populated with Wikipedia entities as nodes, wikipedia edit history could be used to capture changes to the entity. In MDM, we have access to lineage and provenance data from the logs.

\subsection*{Fairness}

Several use-cases where link prediction is used on people graphs have significant societal impact and require an acute awareness of privacy, fairness and data protection. A dataset with reasonable diversity and fairer representation of underprivileged groups is also a desired requirement in our dataset. \cite{vannur2020data} present methods to measure and sample the dataset. These methods in turn have been based on the AIF 360 toolkit \cite{bellamy2018ai}.

\subsection*{Batch and Online Link Prediction}

The Watch-List Links Prediction is a typical use-case in Master Data Management as shown in Figure \ref{fig:watchlist_link_prediction}. When the master data is initially created or updated in batch, there is a need for batch mode link prediction. Later when the model is deployed, and a new node or an update to nodes and edges is performed, there is a need for online link prediction. Considering an enterprise property graph is likely to be updated quite frequently, the model needs to be frequently re-trained which is expensive or we have to accept the drift in model performance. The inductive approach to learning the node representations in GNNs helps to make the models more robust to unseen data, but the inductive approach will not be sufficient if the distribution of the unseen nodes is different from the graph created in batch mode.

\subsection*{Multiple Data Sources}

Data from a single source is typically not enough to capture real world use-cases. Typically the master data in a company is dervied from different sources. The ability to distinguish between different people becomes especially important to avoid wrongful predictions in sensitive use-cases.

\subsection*{Factsheets}

We need a way to capture the characteristics of the datasets and assumptions made during the training process. Similar to Factsheets described in \cite{arnold2019factsheets}, we need a way to present this information to developers who might use the solution.

\subsection*{Desired Dataset Characteristics}

\begin{table}[!htb]
    \begin{center}
    \begin{tabular}{p{8cm}}
    \hline
    \vspace{0.1cm}
    \textbf{Desired Dataset Characteristics} \\
    \vspace{0.1cm} \\
    \hline
    \vspace{0.2cm}
    1. Personal data should be there (name, email, address, even SSN if it can be generated artificially). \\
    2. However the personal data should be synthetically generated. Alternatively original data should be anonymized like in MIMIC III dataset. \\
    3. The ratio of nodes and edges should be configurable or typically be 1:5.\\
    4. For non obvious link prediction, the dataset requires several hops. Hence the average path length should be sufficiently long.\\
    5. Entity re-resolution problem (based on updates to the graph) will need a dataset with temporal information. \\
    6. People can have different names (legal name, business name etc), addresses (permanent, business, contact etc), phone numbers. \\
    7. The number of attributes in enterprise graphs on people are about 50. Including the timestamp and lineage data this can go up to 145 attributes. \\
    8. The datasets should have millions of rows to adequately represent the real world datasets. \\
    9. Ids are weighted higher than other columns in Master Data Management. Hence we need identifier data. \\
    10. Multiple similar subgraphs should be available. This is to train and test on different graphs.\\
    \vspace{0.2cm} \\
    \hline
    \end{tabular}
    \caption{Dataset Characteristics for MDM Link Prediction}
    \label{tab:dataset_requirements}
    \end{center}
\end{table}

\label{characteristics}
In Table \ref{tab:dataset_requirements}, we summarize the desired characteristics in a dataset meant for Watch-List Link prediction.

\section{Master Data Management}
\label{mdm_dataset}

In this section, we describe the Master Data Management solution offered by IBM Infosphere MDM by generating a dataset for Link Prediction from scratch.

\subsection{Motivation}
\label{falsefp}
Wikidata, Wikipedia and crowd sourced data in general have a ‘false’ False Positives problem. If we train a model on such crowd sourced data and the model predicts new links, we will not know if the predicted links are missing links correctly predicted by the model or false positives. Manual evaluation of the correctness of these models is not possible at scale. Hence we need automated, even if approximate methods to verify the links proposed by these models.

When we have an incomplete graph, and the link prediction model predicts a new link not present in the original graph, it is not obvious if the predicted link is a false positive or new link that we wish to predict. Graphs created from Wikipedia and DBPedia especially suffer from this problem, because the Wikipedia pages and relations may not have been created by volunteers. Even in enterprise graphs, the new links predicted may have to be verified by some process, to be not considered as false positives.


To explain this 'false' False Positives problem, we present the results of our Link Prediction experiments on the UDBMS dataset. This dataset is based on the person data in DBPedia. It has 510733 nodes, but only 28945 edges. While Link Prediction is likely to predict lots of missing edges, it'll be hard to distinguish correctly predicted missing links and false positive links.

\begin{table}[!htb]
    \begin{center}
    \begin{tabular}{lcc}
    \hline
    \textbf{} & \textbf{Large Subgraphs}& \textbf{All Subgraphs} \\
    \hline
    AUC*  & 0.5615 & 0.3888  \\
    Accuracy* & 0.4286 & 0.4039   \\
    Positive Samples Accuracy  & 0.7360 & 0.4460  \\
    Positive Predictions \\on Negative Samples  & 0.8709 & 0.5300   \\
    \hline
    \end{tabular}
    \caption{Link Prediction on UDBMS Dataset}
    \label{tab:linkprediction}
    \end{center}
\end{table}

\begin{figure}[!htb]
    \includegraphics[width=\columnwidth]{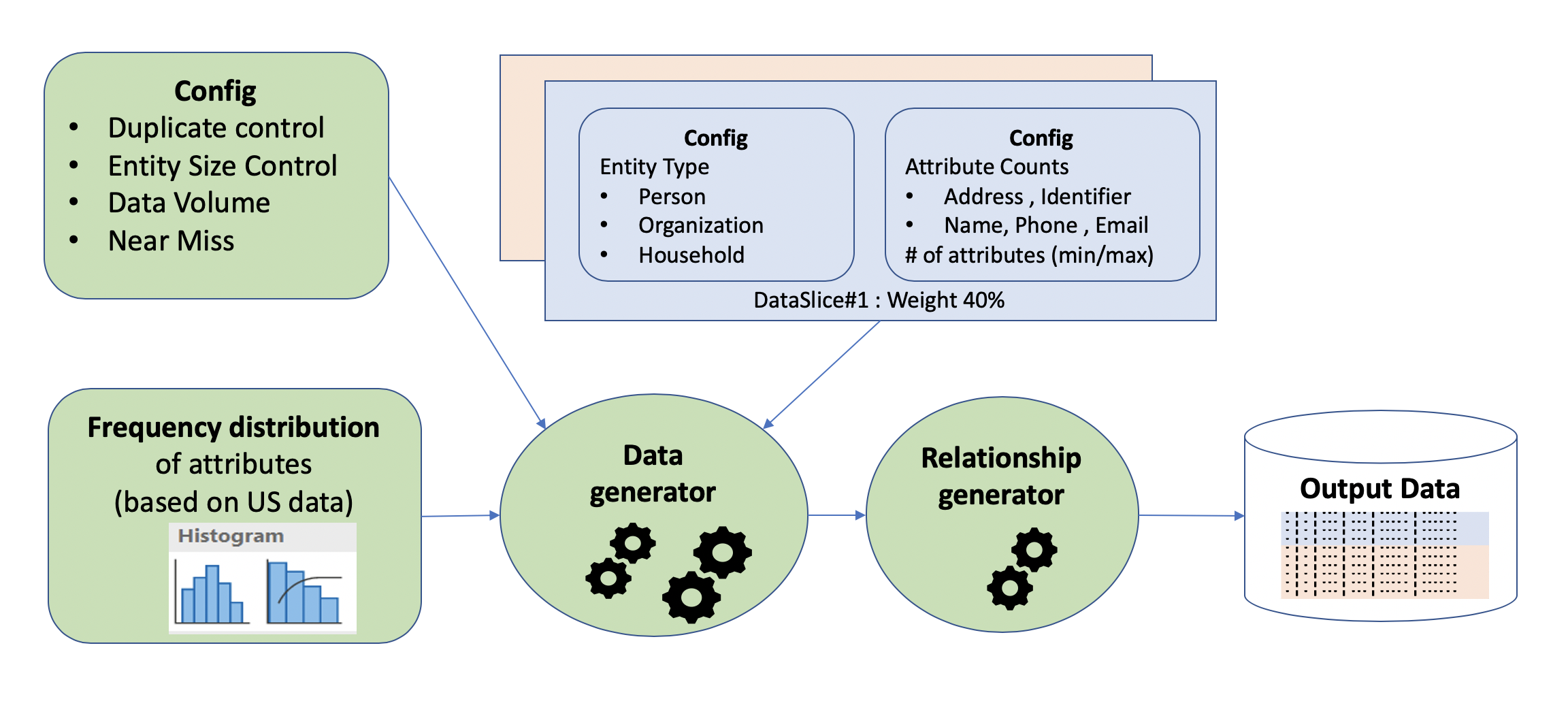}
    \caption{Data Generation using IBM Infosphere MDM}
    \label{fig:data_generator}
\end{figure}

As can be seen in Table \ref{tab:linkprediction}, the AUC and Accuracy scores do not account for the incomplete graph. They assume the negative samples to be not linked, because those nodes are not linked to each other in the dataset, but this could be just because of the incompleteness of the crowd sourced DBPedia content. As model predicts links on these negative samples, the AUC* and Accuracy* will suffer. One way to handle this is to just report the positive samples accuracy - the ability of the model to predict the links that were removed by us for testing. We can also report the positive predictions on the negative samples. In this UDBMS dataset, there seems to be substantial amount of new links predicted which were not present in the original dataset. This could either mean, the model has low precision or it could mean the underlying graph is incomplete. Considering there are only around 28k edges for 510k nodes, significant number of edges might be missing in this dataset. This is one of the prime motivations to create a new dataset for our experiments.

In the rest of this section, we'll present our method to create an internal IBM dataset, which we call as the MDM Bootcamp Dataset. This also helps understand Entity Resolution in Master Data Management. Typically, data about people, organization or other types of entities in the Master Data of a company come from multiple sources before they are combined to form the master data. Hence, we have created this dataset by resolving entities from 3 different data sources in different formats namely, unstructured data (text), semi-structured (json) data and structured data (DB2 tables).

\subsection{Entity Extraction from Text}

\cite{Zitnik2017} introduced the Protein Protein Interaction dataset that has 20 subgraphs each representing a human tissue. Having similar subgraphs presents a number of advantages over other graphs.

We start with the idea of creating a property graph of people consisting of multiple similar subgraphs albeit with different sizes. This will allow us to train and infer on different subgraphs without affecting the connected-ness and hence node representation in the original data. We however leave experiments on this for future work, and train and infer from all the subgraphs as done in \cite{ying2019gnn}.

Hence, we first create a knowledge graph from unstructured text data. We used the automatic Knowledge Base Population (KBP) method that we have described in \cite{vannur2020data}. In addition to neural models for various tasks in KBP, we have introduced a rule based augmentation technique to increase the diversity of the dataset. Given the ethical considerations that we have described in Section \ref{ethical_considerations}, having the ability to control the diversity in personal data becomes important.

\subsection{Scaling and Automation with Data Generator}
\label{data_generator}

\begin{figure*}[htb]
\begin{subfigure}{0.48\textwidth}
    \includegraphics[width=\columnwidth]{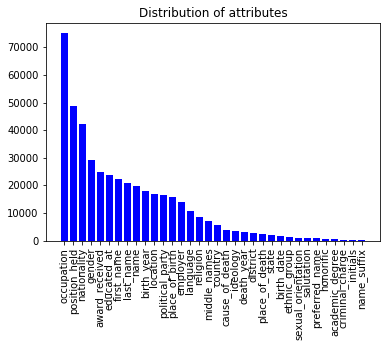}
    \caption{Distribution of node attributes}
    \label{fig:node_attributes_distribution}
\end{subfigure}
\begin{subfigure}{0.48\textwidth}
    \includegraphics[width=\columnwidth]{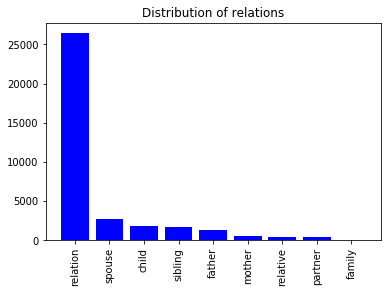}
    \caption{Distribution of relation types}
    \label{fig:relation_distribution}
\end{subfigure}
\end{figure*}

\begin{figure}
    \includegraphics[height=10cm, width=\columnwidth]{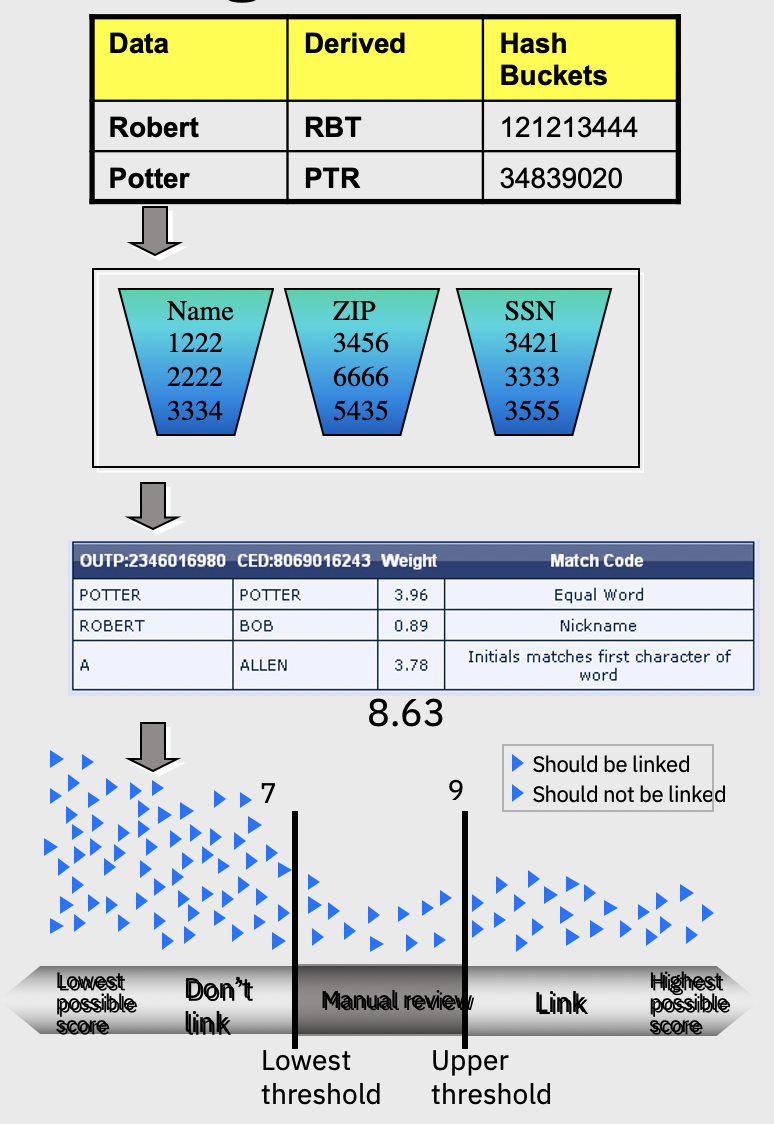}
    \caption{Probabilistic Matching Engine in MDM}
    \label{fig:pme_algorithm}
\end{figure}

In the context of MDM, an entity is quite often a person or an organisation. The key attributes that define a person entity include fields like first name, last name, gender, date of birth, residential address, phone number, identifiers like SSN number, passport number etc. In addition there are fields like references or links to other entities which could be other person entity, an Org entity or a contract entity.

For the purpose of testing functional correctness, performance evaluations, evaluation and training of Deep Learning/AI models,  we need real looking datasets. Obviously, such dataset should not include real life data for the reasons like privacy and confidentiality, which we have utmost respect for. This is also a legal requirement in many countries. Hence we needed a way to generate the real-looking synthetic data.

We have developed an in-house tool called the Data Generator shown in Figure \ref{fig:data_generator} that provides data as per requirements.  The demographic data is generated typically of the distribution of the US population. The frequency distribution of various tokens are taken into account to simulate the real life scenario. We introduce near-misses in attributes to simulate typos. The synthetic data also contains predefined number of suspected duplicates which are created with partial set of attributes matching with that of original entity such that the size of entities distributed as per Zipf's law.

In addition to the attributes for an entity, the synthetic data also contains stated relationships between entities. The data provides the degree of separation between given entities, which can be used to evaluate the capability of the Deep Learning/AI model to find remotely-related entities.

\subsection{Entity Resolution using MDM}

Entity Resolution is the core capability of the IBM Infosphere MDM product. So it is a natural choice for us to resolve the entities present in the 3 data sources. \cite{oberhofer2014beyond} describes the Probabilistic Matching Engine that is at the core of the Entity Matching solution. As shown in Figure \ref{fig:pme_algorithm}, a number of heuristics have been developed over years, for name, addresses, phone numbers, identification numbers that are typical node attributes in Enterprise graphs. From finding edit distance, to complex statistical models, each attribute is handled differently. This approach could help design interpretability solutions like in Neural Additive Models \cite{agarwal2020neural}, which we hope to explore in a later work.

The core challenge however is in performing entity matching at scale and low latency. A number of bucketing techniques have been developed to address this requirement. The output of the engine can be consumed directly using an Auto Link threshold or reviewed manually. Data Stewards optimize the parameters and deploy models customized for different workloads.

\begin{figure}[htb]
    \includegraphics[width=\columnwidth]{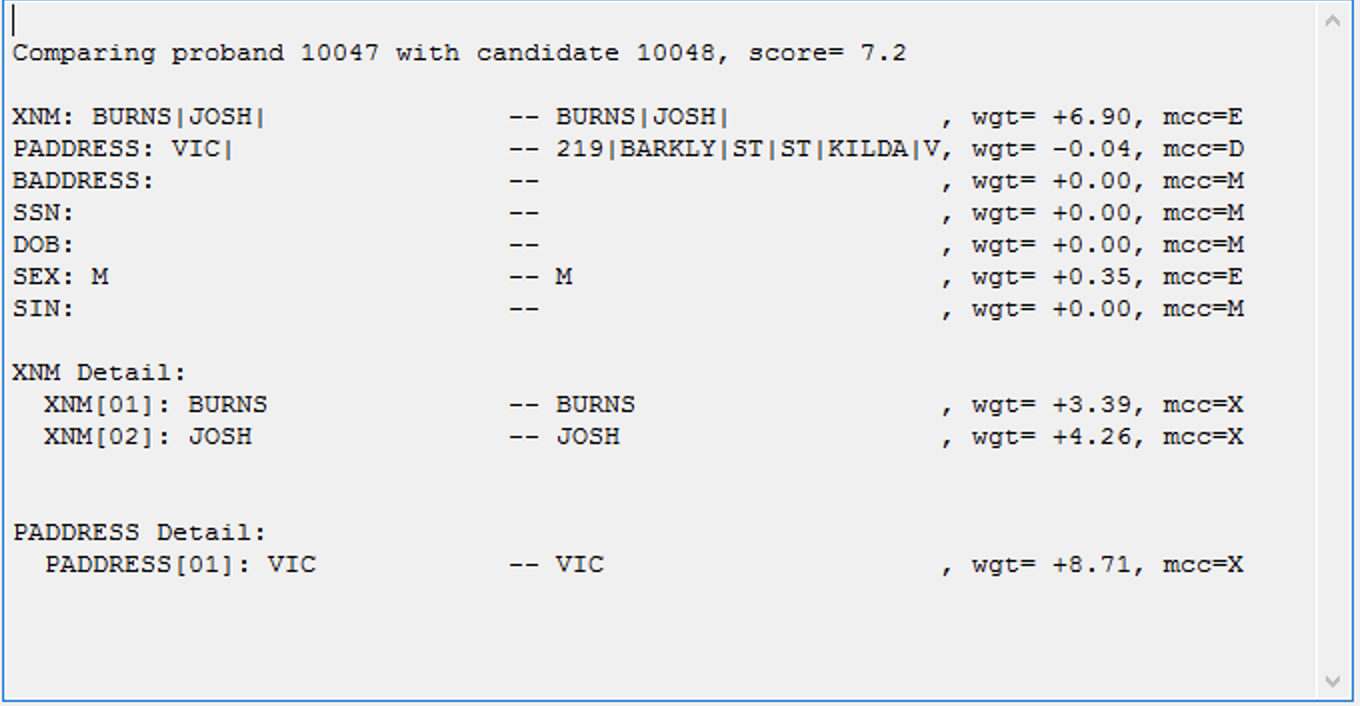}
    \caption{Inspector tool for Data Stewards to inspect and opitimize parameters for Entity Resolution}
    \label{fig:same_entity}
\end{figure}

Applying the above PME algorithm on our example data, we get the score shown in Figure \ref{fig:same_entity}.

\subsection{Anonymization}

Given the sensitivity of the dataset created from unstructured, semi-structured and structured data containing personal data entities, we need to anonymize the data before it can be used to train models. However, anonymizing all the entities could lead to the models failing to predict links. We solve this problem by adopting a strategy similar to MIMIC III dataset \cite{johnson2016mimic}. We use our fine grained entity classification work \cite{dasgupta2018fine} to detect entities and shift them appropriately.

\subsection{Diversity and Fairness}

Compared to the Dataset extracted from just unstructured text text, we observe that the merged dataset obtained from three sources is much richer. As shown in Figure \ref{fig:node_attributes_distribution} and Figure \ref{fig:relation_distribution}, we have generated a medium sized graph, but it's trivial to scale the graph using methods described in Section \ref{data_generator}.

AIF360 \cite{bellamy2018ai} and similar toolkits for detecting bias in tabular data, predominantly use counter factual examples to perturb data to observe if the target variable changes because of the perturbation. We have used the techniques mentioned in \cite{vannur2020data} to augment data such that the diversity increases in protected variables. We can also use the methods described in that work to ensure that downstream applications like Link Prediction do not depend on protected variables like gender, ethnicity etc.

\begin{table}[!htb]
    \begin{center}
    \begin{tabular}{llll}
        \textbf{}           & \textbf{UDBMS}            & \textbf{MDM Bootcamp} \\
        \hline
        \textbf{Persons}         & 510733 & 20912 \\
        \textbf{Links} & 28945 & 67564 \\
        \textbf{Attributes} & 38    & 70 \\
        \textbf{Relations}  & 10   & 9 
    \end{tabular}
    \caption{Statistics on personal data annotations. We use data augmentation on our MDM Bootcamp dataset to closely resemble enterprise datasets.}
    \label{tab:datasets}
    \end{center}
\end{table}

\begin{figure*}[htb]
\begin{subfigure}{0.5\textwidth}
    \centering
    \includegraphics[width=\columnwidth]{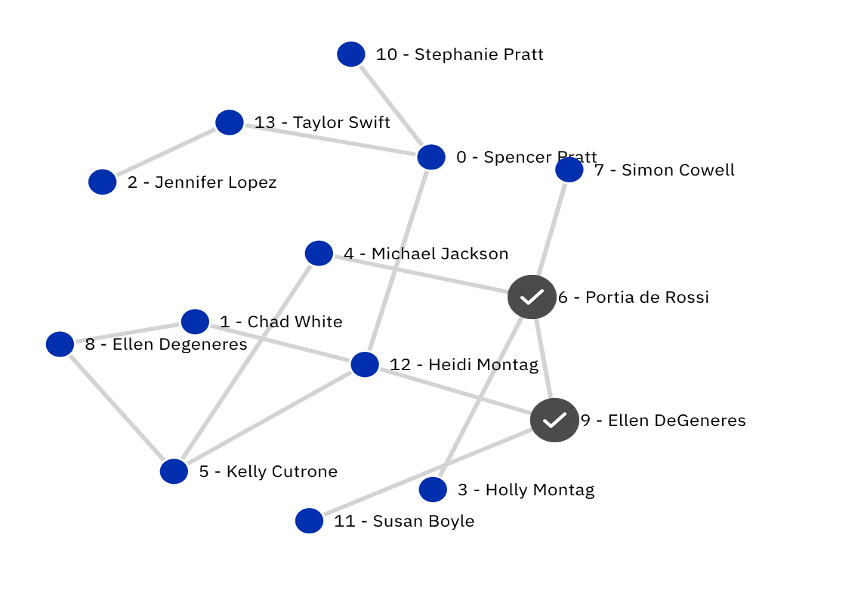}
    \caption{A subgraph with links predicted links}
    \label{fig:example_graph}
\end{subfigure}
\begin{subfigure}{0.5\textwidth}
    \centering
    \includegraphics[width=\columnwidth]{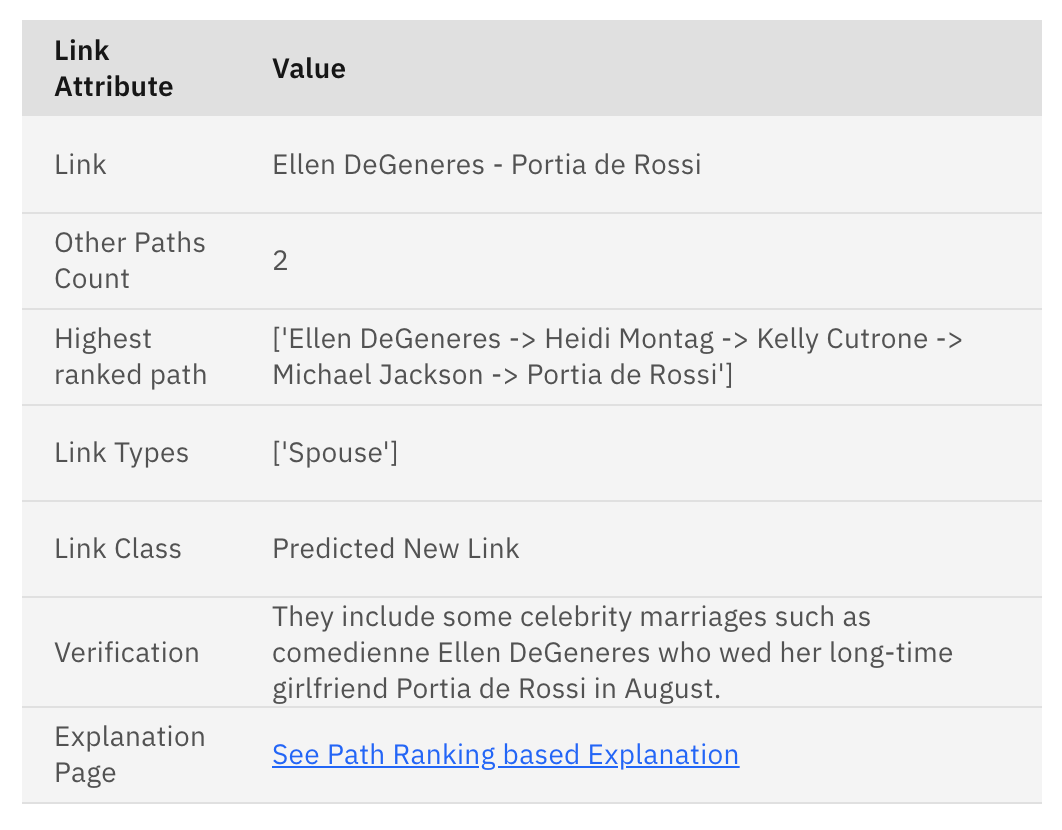}
    \caption{Explanations for a predicted link}
    \label{fig:verifiable_text}
\end{subfigure}
\end{figure*}
\begin{figure}[htb]
    \centering
    \includegraphics[width=\columnwidth]{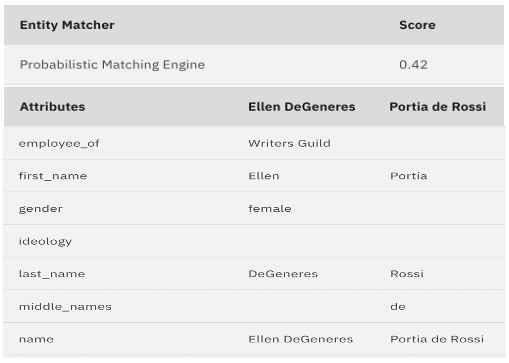}
    \caption{Comparison of nodes involved in a predicted link}
    \label{fig:node_comparison}
\end{figure}

\section{Link Prediction Implementation}
\label{experiments}

We begin by explaining our experimental setup. We use the framework provided by \cite{you2019position}, which in turn uses pytorch \cite{paszke2019pytorch} and more specifically pytorch geometric \cite{fey2019fast}. One of the pre-processing steps we do is finding the all pairs shortest paths calculation using appropriate approximations. As we'll see in Section \ref{explainability}, this pre-processing step comes in handy to explain the links as well.

We train models using GCN and PGNN on the UDBMS Dataset and the MDM Bootcamp Dataset that we have described in Section \ref{mdm_dataset}. Following the procedure in \cite{you2019position}, we choose only connected components with atleast 10 nodes for our experiments. A positive sample is created by randomly choosing 10\% of the links. For the negative sample, we use one of the nodes involved in the positive samples and pick a random unconnected node as the other node. The number of negative samples is same as that of the positive samples. We discuss hard negative samples in our future work section. Our batch size is typically 8 subgraphs and for PGNN, we use 64 anchor nodes.

\begin{table}[!htb]
    \begin{center}
    \begin{tabular}{cccc}
    \hline
    \textbf{Dataset} & \textbf{Model} & \textbf{ROC AUC} & \textbf{Std. Dev.} \\
    \hline
    {\textbf{UDBMS}}  &   \textbf{GCN} & 0.4689 & 0.0280   \\
                    &   \textbf{P-GNN} & 0.6456 & 0.0185   \\
                        \cline{2-4}
    {\textbf{MDM Dataset}}  &   \textbf{GCN} & 0.4047  & 0.09184  \\
                    &   \textbf{P-GNN} & 0.6473 & 0.02116   \\
                                            \cline{2-4}
    \hline
    \end{tabular}
    \caption{Comparison of Link Prediction performance on different People Datasets}
    \label{tab:comparison}
    \end{center}
\end{table}

\begin{figure}[htb]
    \centering
    \includegraphics[width=\columnwidth]{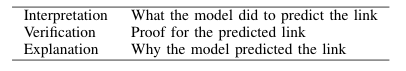}
    \label{fig:legend}
    \caption{We group explainability solutions into 3 categories}
\end{figure}

\begin{figure*}[htb]
    \centering
    \includegraphics[width=\linewidth]{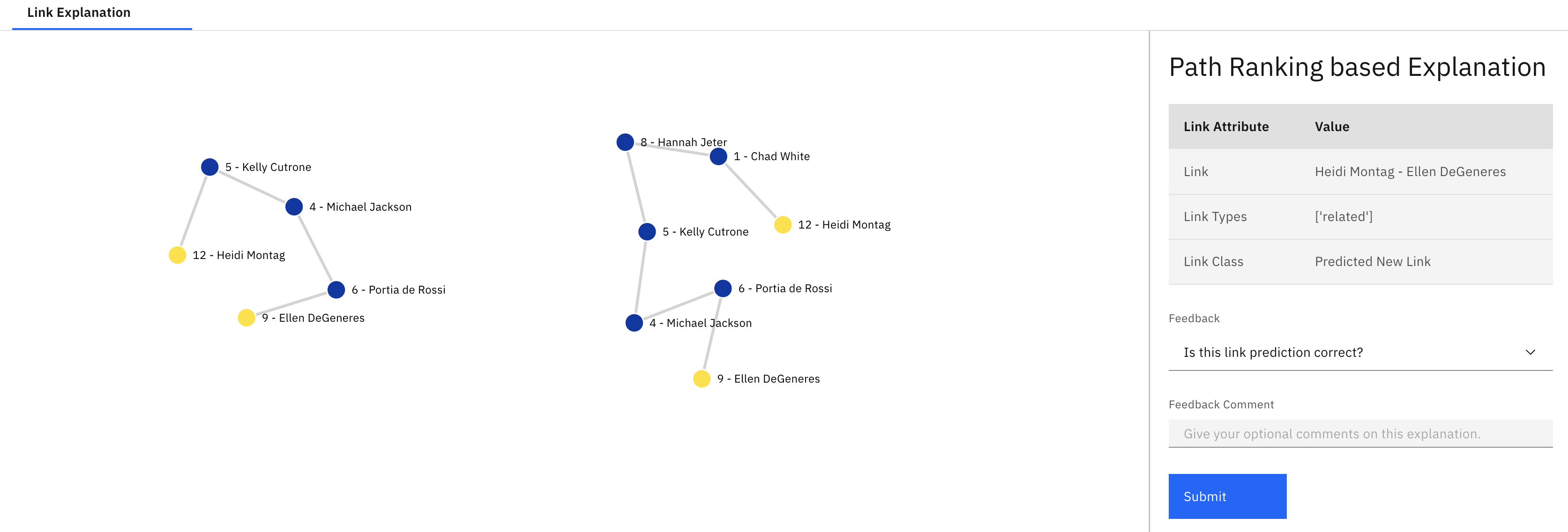}
    \caption{Path Ranking based explanation for Link Prediction. Data Stewards can explore the already existing paths (if any) and understand the predicted link. They can also provide feedback.}
    \label{fig:pbe}
\end{figure*}

As can be seen in Table \ref{tab:comparison}, P-GNN outperforms GCN in both our datasets. The performance obtained by the better model is also less than what we desire. There are a number of options available to try and improve the performance of the models. However, in this work, we want to focus on how such a system can be deployed in production under the professional oversight of Data Stewards.

\subsection{Explainability}
\label{explainability}

In typical enterprise deployments, there will usually be a Data Steward who maintains the master data. It is desired that the models provide explanations so that the Data Stewards can understand why the model is predicting a link. While Data Stewards can also use the system to detect errors, our goal is not to use precious human resources for error detection. Instead, we want the Data Stewards to draw valuable insights into the model's learning process and also understand it's limitations.

In this section, we discuss several explainability techniques that could be used for explaining the links predicted by the Graph Neural Network models in Section \ref{experiments}.

While a number of above such solutions for explaining neural models exist, our focus is been on human understandable explanations for graph neural networks. We explored different techniques in this area and categorized them into three.


\subsection*{Verification}
Verification is the proof for the predicted link. This can be used to understand links already present in the graph and also try to validate new links predicted by our models. Some of the earlier error detection work involved external information to verify the predicted link. \cite{thorne2018fever} describes the FEVER shared task for facts extraction and verification.

\subsection*{Interpretation}
\cite{ribeiro2018anchors} introduced the idea of Anchors as Explanations, which builds upon their earlier LIME solution \cite{ribeiro2016should}. The key idea here is to show only few important features (anchors) rather than showing the pros and cons of several or all features. \cite{huang2020graphlime} introduced GraphLIME, which as the name indicates is a version of LIME for Graph Neural Networks. They compared their solution GNN Explainer \cite{ying2019gnnexplainer} which follows a subgraph approach to explain predicted links.

\subsection*{Explanation}

\cite{elton2020self} presents the Self-Explainable AI idea where a single model is used to both predict and provide parameters that can be used to explain the prediction. This is similar to a multi-task learning setting. Mutual Information is used to measure the faithfulness of the explanation model to the diagnostic model. \cite{sun2019infograph} also uses Mutual Information to measure the closeness of supervised and unsupervised models.

\subsection{Link Verification with External Information}

One of the desirable properties of the dataset that we set out to create is the ability to verify the links present in the dataset. \cite{thorne2018fever} introduced the FEVER dataset for fact verification. We follow a similar method in our work to generate verifiable text that a human annotator can use to decide if the predicted link is valid or otherwise.

While in this dataset, we use text from unstructured pages, we can also include links from other sources namely organization charts, logs, emails and other enterprise documents to generate such verification text. The verifiable text for our example is as shown in Figure \ref{fig:verifiable_text}.

For links that are present in training data, it'll be easy to fetch the verifying text. During inference on unseen data, we could using an information retrieval system, typically a search index to query for text and other details that mention the entities that are predicted to be linked by the Graph Neural Network.

\subsection{Path Ranking based Link Explanation}

Our next explainability solution is inspired by ideas in Error detection in Knowledge Graphs. In particular, we use the PaTyBRED approach described in \cite{melo2017detection} and \cite{meloautomatic}.

Path Ranking Algorithm (PRA) was introduced by \cite{lao2011random} was based on the probability of ending up at an Object o, if we perform random walks on a graph starting at Subject s and Relation r. \cite{gardner-mitchell-2015-efficient} replaced the path probability with binary values. PaTyBRED further simplified the process by using a Random Forest instead of logistic regression and introducing a k-selection method.

As shown in Figure \ref{fig:pbe}, we use the ranking function in the PaTyBRED algorithm to find a path among multiple already existing paths between two nodes. The idea here is that the information contained in that path is more useful than other paths to understand the predicted link. This idea is somewhat similar to the GNN Explainer idea of exploring the subgraph around the predicted link, except that we prefer to use an independent algorithm to rank all the paths rather than subgraph around the two nodes.

\begin{figure*}[!htb]
    \centering
    \includegraphics[height=22cm, width=0.9\linewidth]{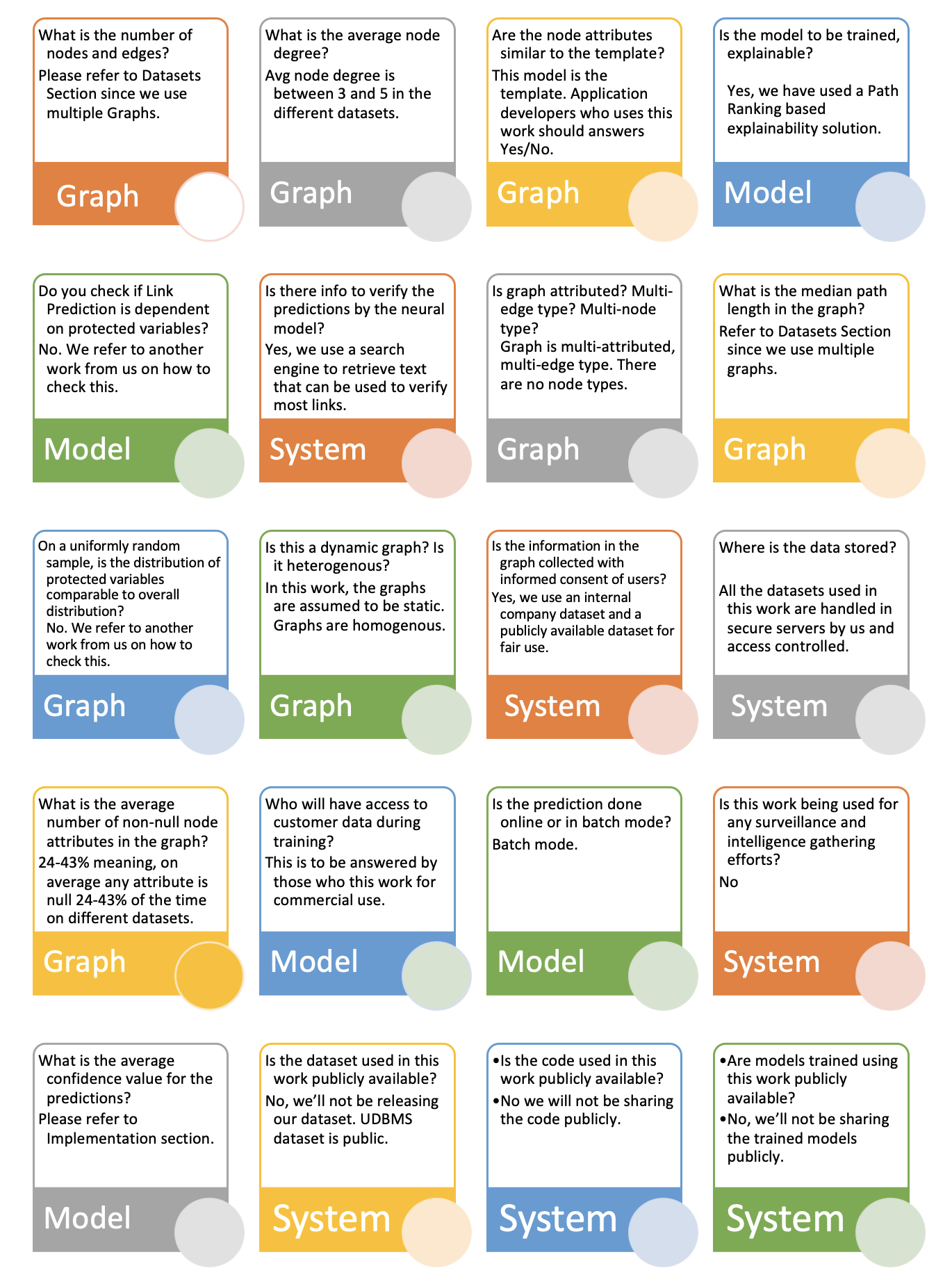}
    \caption{An example Graphsheet to ensure the models are trained similar to the original environment.}
    \label{fig:graphsheets}
\end{figure*}

\subsection{Graphsheets}

After having used the MDM Bootcamp Dataset for training models and evaluating Link Prediction for the Watch List use-case, it is important to ensure that downstream applications and model builders follow similar practices for fairness, privacy, data protection and ethical reasons. Towards this goal, we introduce Graph Sheets. Based on Factsheets \cite{arnold2019factsheets}, Graph Sheets include facts about the graph being studied and a number of FAQ types questions and answers that model developers have to consider before training Link Prediction models on their own datasets. As shown in Figure \ref{fig:graphsheets}, we provide information regarding the model, graph and the systems used in training Graph Neural Models on the MDM Bootcamp dataset.

\section{Future Work}

\subsection*{Entity Re-resolution}

In this work, we have treated the graph to be static while conducting experiments and predicting links. While batch mode link prediction is still the predominant use-case, there have been lot of work in dynamic graphs. We plan to address updates to the graph bases on recent works like EvolveGCN \cite{pareja2020evolvegcn} and TGN \cite{rossi2020temporal}.

\subsection*{Hard Negative Sampling}

Hard negative samples are entities that are closer to our entity of interest in the embedding space, but are still not related to our entity. For example, people having same first names and living in a city are not necessarily related and hence the model should not predict only based on those node attributes. PinSAGE work described in \cite{you2019position} uses images ranked 2000 to 5000 by GraphSAGE as hard negative samples.

\subsection*{Training on similar subgraphs}

Typical Link Prediction experiments tend to split the training data into train, test and validation sets by removing some existing links. Training models by removing some existing links, could potentially affect the performance of the models. We can try training on several subgraphs, and testing on a hold out subgraph. This can be for k-fold validation on the subgraphs or for inferring links on a new subgraph with similar characteristics.

\section{Conclusion}
We introduced a solution for the Link Prediction task on Property Graphs in IBM's Master Data Management product. We presented a number of motivating experiments and use-cases, and describe the special care we take to address fairness, data protection, privacy and AI Ethics. We then presented two human understandable solutions to explain the links prediction by Graph Neural Networks, search based retrieval of verification text and path ranking based explanations.

\section*{Acknowledgements}

We thank Sameep Mehta, Scott Schumacher, Marcus Boon, Berthold Reinwald, Xiao Qin, Paolo Scotton, Thomas Gschwind, Christoph Miksovic, Jim O'Neil, Martin Oberhofer, Lars Bremmer, Manfred Overs, Avirup Saha, Lingraj S Vannur, and Philip Mueller for the discussions and feedback.

\bibliographystyle{IEEEtran}
\bibliography{IEEEabrv,sample-base.bib}

\end{document}